\documentclass[a4paper,11pt]{article}
\pdfoutput=1 

\usepackage{jcappub} 

\usepackage[T1]{fontenc} 

\title{\boldmath Inverse problem for extragalactic transport of ultra-high energy cosmic rays}

\author{V.S. Ptuskin}
\author{S.I. Rogovaya}
\author{and V.N. Zirakashvili}
\affiliation{Pushkov Institute of Terrestrial Magnetism,
Ionosphere and Radio Wave Propagation of the Russian Academy of
Sciences (IZMIRAN),\\Troitsk, Moscow 142190, Russia}

\emailAdd{vptuskin@izmiran.ru}
\emailAdd{rogovaya@izmiran.ru}
\emailAdd{zirak@izmiran.ru}

\abstract{The energy spectra and composition of ultra-high energy
cosmic rays are changing in a course of propagation in the
expanding Universe filled with background radiation. We developed
a numerical code for solution of inverse problem for cosmic-ray
transport equations that allows the determination of average source
spectra of different nuclei from the cosmic ray spectra observed at the
Earth. Employing this approach, the injection spectra of protons
and Iron nuclei in extragalactic sources are found assuming that
only these species are accelerated at the source. The data from the
Auger experiment and the combined data from the Telescope Array + HiRes
experiments are used to illustrate the method.}

\begin{document}
\maketitle
\flushbottom

\section{Introduction}
\label{sec:intro}

The origin of cosmic rays with energies $E>10^{18}$ eV remains a
key problem of cosmic ray astrophysics. The observed suppression
of cosmic ray flux at energies above $\sim 5\times 10^{19}$ eV
seems confirm the presence of the GZK cutoff predicted in
\cite{Greisen66, ZatsKuzm66} although the suppression due to the
acceleration limits in cosmic ray sources can not be excluded
\cite{AllardRev, AloisoBerBlasi13}. The occurrence of the GZK
suppression and the high isotropy of the highest energy cosmic
rays are indicative of their extragalactic origin. The list of
potential sources which could give the observed cosmic ray flux
includes active galactic nuclei, gamma-ray bursts, fast spinning
newborn pulsars, interacting galaxies, large-scale structure
formation shocks and some other objects, see reviews
\cite{Olinto10,Lemoine12,BlasiRev14} and references therein.

The present knowledge about the highest energy cosmic rays was
mainly acquired from the High Resolution Fly's Eye Experiment (HiRes),
Pierre Auger Observatory (Auger), Telescope Array experiment (TA),
and from the Yakutsk complex EAS array,
see \cite{Olinto10,Troits13,Watson14}. The mass composition of
these cosmic rays remains uncertain. The interpretation of HiRes
and TA data favors predominantly proton composition at energies
$10^{18}$ to $5\times 10^{19}$ eV, whereas the Auger data indicate that
the cosmic ray composition is becoming heavier with energies
changing from predominantly proton at $10^{18}$ eV to more heavy
and approaching Iron composition at about $5\times
10^{19}$ eV. The mass composition interpretation of the measured
quantities depends on the assumed hadronic model of particle interactions
which is based on not well determined extrapolation of physics from lower
energies.

The energy spectrum in extragalactic sources is commonly
determined by the trial-and-error method when one makes the
calculations of the expected at the Earth cosmic ray intensity
assuming some shape of the source energy spectrum and the source
composition. The calculations follow cosmic ray propagation from
the source to the observer, e. g. \cite{Allard}. The standard
assumption is that the source spectrum is a power law on magnetic
rigidity up to some maximum rigidity.

In the present work we show how to inverse the procedure and
calculate the source function starting from the observed at the
Earth spectrum without ad hoc assumptions about the shape of
source spectrum. Simple cases of the source composition that includes
protons and Iron nuclei are considered and the analytical approximations
of the data from Auger and TA+HiRes experiments are used.

\section{Solution of inverse problem for a system of cosmic-ray transport equations}
\label{sec:equation}

We use the following transport equation for cosmic ray protons and
nuclei in the expanding Universe filled with the background
electromagnetic radiation (see \cite{ASR} for details):

\begin{eqnarray}
    -H(z)(1+z)\frac{\partial}{\partial
    z}\left(\frac{F(A,\varepsilon,z)}{(1+z)^{3}}\right)- \; \nonumber \\
    -\frac{\partial}{\partial\varepsilon}
    \left(\varepsilon\left(\frac{H(z)}{(1+z)^{3}}+
  \frac{1}{\tau(A,\varepsilon,z)}\right)F(A,\varepsilon,z)
  \right)+
   \nu(A,\varepsilon,z)F(A,\varepsilon,z) \nonumber \\
  = \sum_{i=1,2...}\nu(A+i\rightarrow
  A,\varepsilon,z)F(A+i,\varepsilon,z)+q(A,\varepsilon)(1+z)^{m}.\;
  \label{eq:transport}
\end{eqnarray}

The system of eqs. (\ref{eq:transport}) for all kinds of nuclei with
different mass numbers $A$ from Iron to Hydrogen should be solved
simultaneously. The energy per nucleon $\varepsilon=E/A$ is used
here because it is approximately conserved in a process of nuclear
photodisintegration, $F(A,\varepsilon,z)$ is the corresponding
cosmic-ray distribution function, $z$ is the redshift, $q(A,\varepsilon)$ is the
density of cosmic-ray sources at the present epoch $z=0$, $m$
characterizes the source evolution (the evolution is absent for
$m=0$), $\tau(A,\varepsilon,z)$ is the characteristic time of
energy loss by the production of $e^{-}e^{+}$ pairs and pions,
$\nu(A,\varepsilon,z)$ is the frequency of nuclear
photodisintegration, the sum in the right side of eq. (\ref{eq:transport})
describes the contribution of secondary nuclei produced by the
photodisintegration of heavier nuclei,
$H(z)=H_{0}((1+z)^{3}\Omega_{m}+\Omega_{\Lambda})^{1/2}$ is the
Hubble parameter in a flat universe with the matter density
$\Omega_{m}(=0.3)$ and the $\Lambda$-term
$\Omega_{\Lambda}(=0.7)$.

The numerical solution of cosmic-ray transport equations
follows the finite differences method. The variables are the
redshift $z$ and $\log(E/A)$. The maximum value $z_{max} =3$ is assumed in our calculations.

A comprehensive analysis of cosmic ray propagation in the
intergalactic space was presented in \cite{Berez06}.

Equations \ref{eq:transport} are valid for an arbitrary regime of cosmic ray
propagation - diffusion, rectilinear motion, or any intermediate
regime. It should be emphasized that an alternative Monte Carlo techniques
were used for treating the propagation and interactions of ultra-high
energy cosmic rays. A representative list of references was given
in \cite{Aloisio13}. Although more
sophisticated, the
Monte Carlo method is in general more time consuming compared to the
solution of equations \ref{eq:transport} by the finite differences method.

The assumption of continuous source distribution is not valid when
particles lose energy at a scale less than the distance between
cosmic ray sources. According to the latest Auger results, the low
bound on cosmic ray sources density is estimated as
$n_{s}\geq (0.06-5)\times10^{-4} \textrm{Mpc}^{-3}$, \cite{AUGER2013}.
The finite distance to the nearest source is
approximately taken into account in our calculations by the cutoff
of the source distribution at $z_{min}\approx 0.48H_{0}d/c << 1$,
so that $q=0$ at $z\leq z_{min}$ ($0.48d$ is the average distance
of an observer to the nearest source if point sources arrange
a cubic lattice with the edge, the distance between sources,
equals to $d$). The statistically uniform source distribution is
assumed at larger redshifts.

The stochastic nature of interactions is not taken into account
in our transport equations. The Monte Carlo modelling that includes this
effect and the analytical
model, which is a simplified version of the model used in our calculations,
were compared in \cite{Hooper08}. A fairly
good agreement of both calculated spatial distributions of secondary
species produced by an isolated source of ultra high-energy cosmic rays
was found. A similar conclusion was made in the work \cite{Taylor11}.
The difference for the Iron source is less than $15$ percent that
satisfies the needs of our work. The ultra-high energy proton spectrum
is affected by fluctuations in the photopion production. The noticeable
effect of fluctuations is expected at energies $E\geq1\times10^{20}$ eV.
The Monte Carlo simulations of the proton propagation at
$n_{s}=10^{-5} \textrm{Mpc}^{-3}$ \cite{Berezinsky2006}
showed not more than $\sim10$ percent difference at $E>6\times10^{19}$ eV
compared to the calculations made in the model with continuous energy loss
and the homogeneous source distribution. These results characterize the
errors in our approximations of cosmic ray transport process.

Let us introduce solution $G(A,\varepsilon;A_{s},\varepsilon_{s})$
of eqs. (\ref{eq:transport}) at $z=0$ for a delta-source
$q(A,\varepsilon)=\delta_{AA_{s}}\delta(\varepsilon-\varepsilon_{s})$.
This source function describes the emission of nuclei with mass
number $A_{s}$ and energy $\varepsilon_{s}$ from cosmic ray
sources distributed over all $z$ up to $z_{\textrm{max}}$.
The general solution of eqs. (\ref{eq:transport}) at the observer location
$z=0$ can now be presented as

\[
F(A,\varepsilon,z=0)=\sum_{A'}\int
d\varepsilon'G(A,\varepsilon;A',\varepsilon')q(A',\varepsilon'),
\]
\begin{equation}
N(A,E,z=0)=A^{-1}\sum _{A'}\int
d\varepsilon'G(A,E/A;A',\varepsilon')q(A',\varepsilon')
\label{eq:green}
\end{equation}
Here $N(A,E,z)$ is the spectrum of nuclei with atomic mass $A$ as the
function of the total energy $E$. The observed
all-particle spectrum is determined by the summation
over all types of nuclei $\sum_{A}N(A,E,z=0)$.

The set of discrete values of particle energies $E_{i}$ and  $\varepsilon_{i}$ is
defined to solve the transport equation numerically. The grid with
constant $\triangle \varepsilon /\varepsilon$ and with $100$
energy bins per decade is used in our calculations. Eq. (\ref{eq:green})
in the discrete form is
\begin{equation}
N_{i}(A,z=0)=A^{-1}\sum_{j,A'}(\triangle
\varepsilon)_{j}G_{ij}(A;A')q_{j}(A'),
\label{eq:greennum}
\end{equation}
where the subscript indexes $i$ and $j$ denote the corresponding
energies $E_{i}$ and $\varepsilon_{j}$. The all particle
spectrum is $\sum_{A}N_{i}(A,z=0)$.

The source term $q_{j}(A)$ for each type of nuclei can be derived
from the system of linear eqs. (\ref{eq:greennum}) if the observed spectra
$N_{i}(A,z=0)$ for all types of nuclei are known. In fact, the
detailed information on the spectra of individual types of nuclei
$N_{i}(A,z=0)$ is usually not available.

If only the all particle spectrum $\sum_{A}N_{i}(A,z=0)$ is known,
the source spectra can be found when the source abundances for
different types of ions are specified. In the simplest case, when
only nuclei with mass number $A=A_{s}$ are accelerated in the
sources, eq. (\ref{eq:greennum}) allows to find the following relation:
\begin{equation}
q_{j}(A_{s})=\sum_{i}(\sum_{A}A^{-1}(\triangle\varepsilon)_{j}G_{ij}(A,A_{s}))^{-1}\times \sum_{A}N_{i}(A,z=0).
\label{eq:qGreen}
\end{equation}
In another physically justified case, the source terms for all
primary nuclei are similar functions of magnetic rigidity
$R = E/Z$, so that $q(A,\varepsilon)=S_{A}Q(A\varepsilon/Z)$
where $S_{A}$ are the normalization coefficients. The last equation
in a matrix form can be presented as
\begin{equation}
q_{j}(A')=\sum_{k} M_{jk}(A')Q_{k},
\label{eq:linear}
\end{equation}
where $\textbf{M}$ is a matrix, which provides the needed
dependence of the source term on rigidity. The relation $Z=A/2$ is
assumed in our calculations for all nuclei heavier than protons.

The formal solution of the inverse problem is now
\begin{equation}
Q_{k}=\sum_{i}(\sum_{A,A',j}A^{-1}(\triangle\varepsilon)_{j}G_{ij}(A,A')M_{jk}(A'))^{-1}\times \sum_{A}N_{i}(A,z=0).
\label{eq:qMGreen}
\end{equation}
Technically, the calculations of inverse matrixes in eqs.
(\ref{eq:qGreen}, \ref{eq:qMGreen}) are straightforward since $G_{ij}$ is the
triangular matrix owing to the monotonic decrease of particle energy
in a course of its transport in the intergalactic medium. However, as it
is known from the analysis of
equations of such kind \cite{Turchin71}, the solutions of the
integral and matrix eqs. (\ref{eq:green}, \ref{eq:greennum}) do not depend continuously
on their right-hand sides. The small errors in the data
$N(A,E,z=0)$ may be greatly amplified in the solution
$q(A,\varepsilon)$, so that the inverse problem is ill-posed. In
essence, the difficulty posed by inverse problems is that we are
obliged to work not with exact $N(A,E,z=0)$ but with its
estimate obtained by measurements and therefore subject to
accidental errors. Additional errors occur because of the approximate
description of cosmic ray propagation. Different regularization procedures can be used
to deal with these problems \cite{Turchin71,Tikhonov77,Lucy94}.
Below we try to work in the region of parameters where the regularization
is not required. In particular, the consideration is limited to not
more than two types of nuclei in the sources, the protons and Iron, and the
Iron-to-proton source ratio is not too low that alleviates the problem.

\section{Approximation of experimental data}
\label{sec:data}

To simplify calculations and damp the spread of data points in the
measured at the Earth cosmic ray spectrum, we use its analytical
approximations.

\begin{figure}[tbp]
\begin{center}
\includegraphics[width=10.0cm]{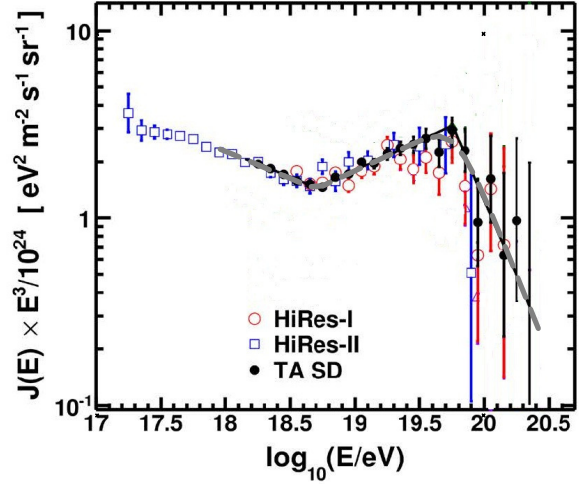}
\end{center}
\caption{Analytical approximations used in the present calculations to describe
TA+HiRes data are shown by dash gray line together with the experimental
data and its straight-line approximation (thin black lines) \cite{TAHiRes13}.}
 \label{TAHiRes13dat}
\end{figure}

The combined results of the TA and HiRes measurements \cite{TAHiRes13}
are presented in figure \ref{TAHiRes13dat}, where the straight lines show the
approximation suggested in \cite{TAHiRes13}. Based on this approximation, we use
the following formula for the observed at the Earth spectrum in our calculations:
\begin{eqnarray}
J\propto E^{-3.283}, E<5.04\times 10^{18} \textrm{eV}; \; \nonumber \\
J\propto E^{-2.685}\times [1+(E/(5.8\times 10^{19}\textrm{eV}))^{1.935}\times \; \nonumber \\
exp(-(7\times 10^{19}\textrm{eV}/E)^{2})]^{-1}, E>5.04\times 10^{18} \textrm{eV}. \;
\label{eq:analytTA}
\end{eqnarray}
The corresponding approximation of the TA+HiRes data is shown by the dash
gray line in figure \ref{TAHiRes13dat}.

\begin{figure}[tbp]
\begin{center}
\includegraphics[width=11.0cm]{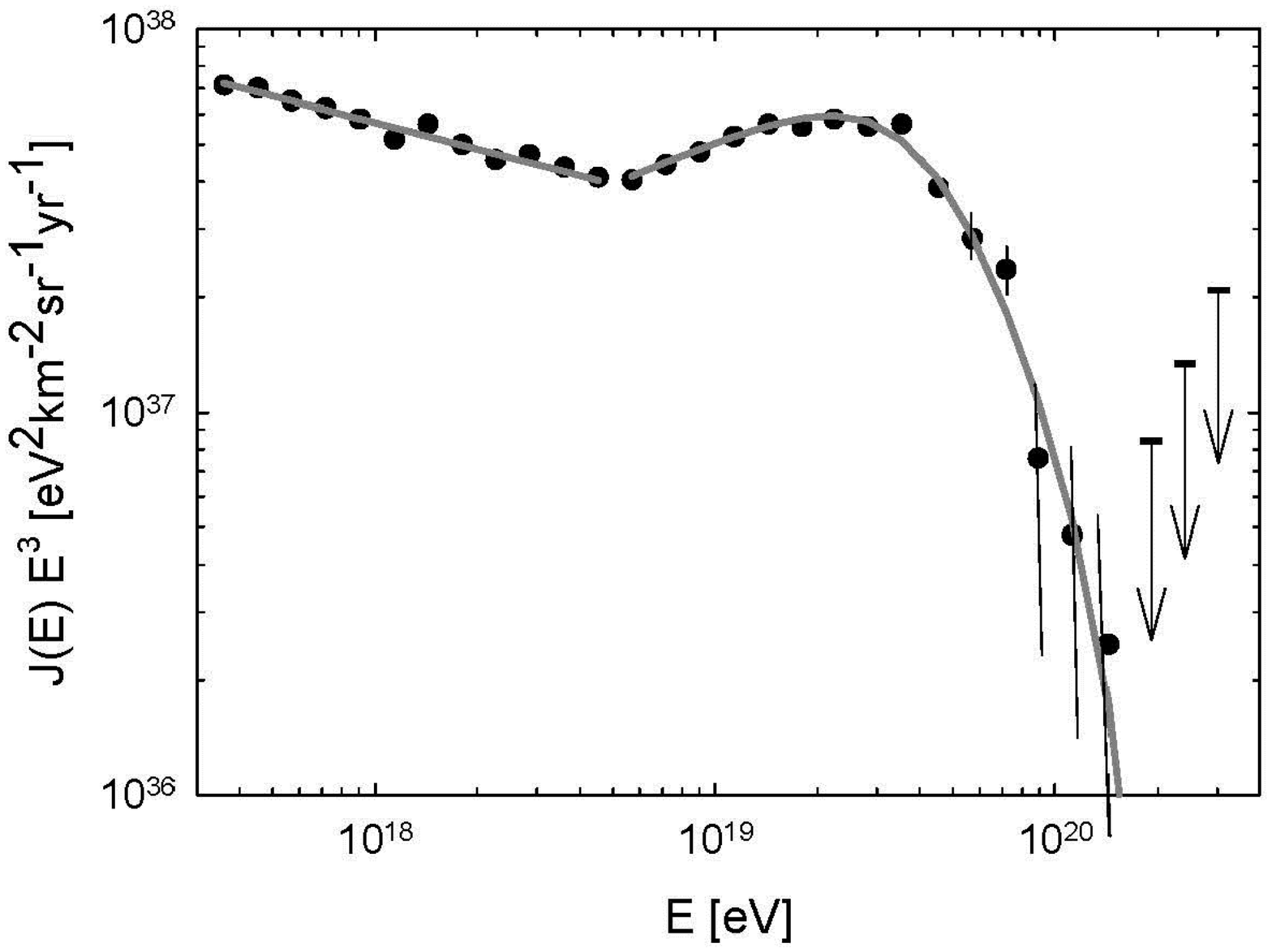}
\end{center}
\caption{Analytical approximations used in the present calculations to describe
the Auger data are shown by solid line together with Auger data \cite{Auger13}.}
 \label{Auger13dat}
\end{figure}

The formula
\begin{eqnarray}
J(E)\propto E^{-3.23}, E < 5\times 10^{18}\textrm{eV}; \; \nonumber \\
J(E)\propto E^{-2.63}\times [1+exp(log(E/10^{19.63}\textrm{eV})/0.15)]^{-1}\times \; \nonumber \\
exp(-(E/(1.5\times 10^{20}eV))^{4}), E>5\times10^{18}\textrm{eV}. \;
\label{eq:analytAuger}
\end{eqnarray}
is used in our calculations to approximate the Auger data  \cite{Auger13}, see
figure \ref{Auger13dat}. This formula is similar to the equation suggested by
the Auger team but contains $exp(-(E/1.5\times 10^{20}eV)^{4})$ factor of cosmic ray flux
suppression at energies $\gtrsim 1.5\times 10^{20}$ eV.

Using the Auger data on energy dependence of the mean logarithm of the
atomic mass number $\langle lnA \rangle$
calculated in the EPOS-LHC model of particle interactions in the
atmosphere \cite{Auger13}, we accept the following approximation
\begin{equation}
\langle lnA \rangle =0.5+4.2\times (E/10^{20}\textrm{eV})^{0.6}
\label{lnA}
\end{equation}
shown by the dash line in figure \ref{composition} below.

\section{Results of Calculations}
\label{sec:calculations}

\begin{figure}[tbp]
\begin{center}
\includegraphics[width=10.0cm]{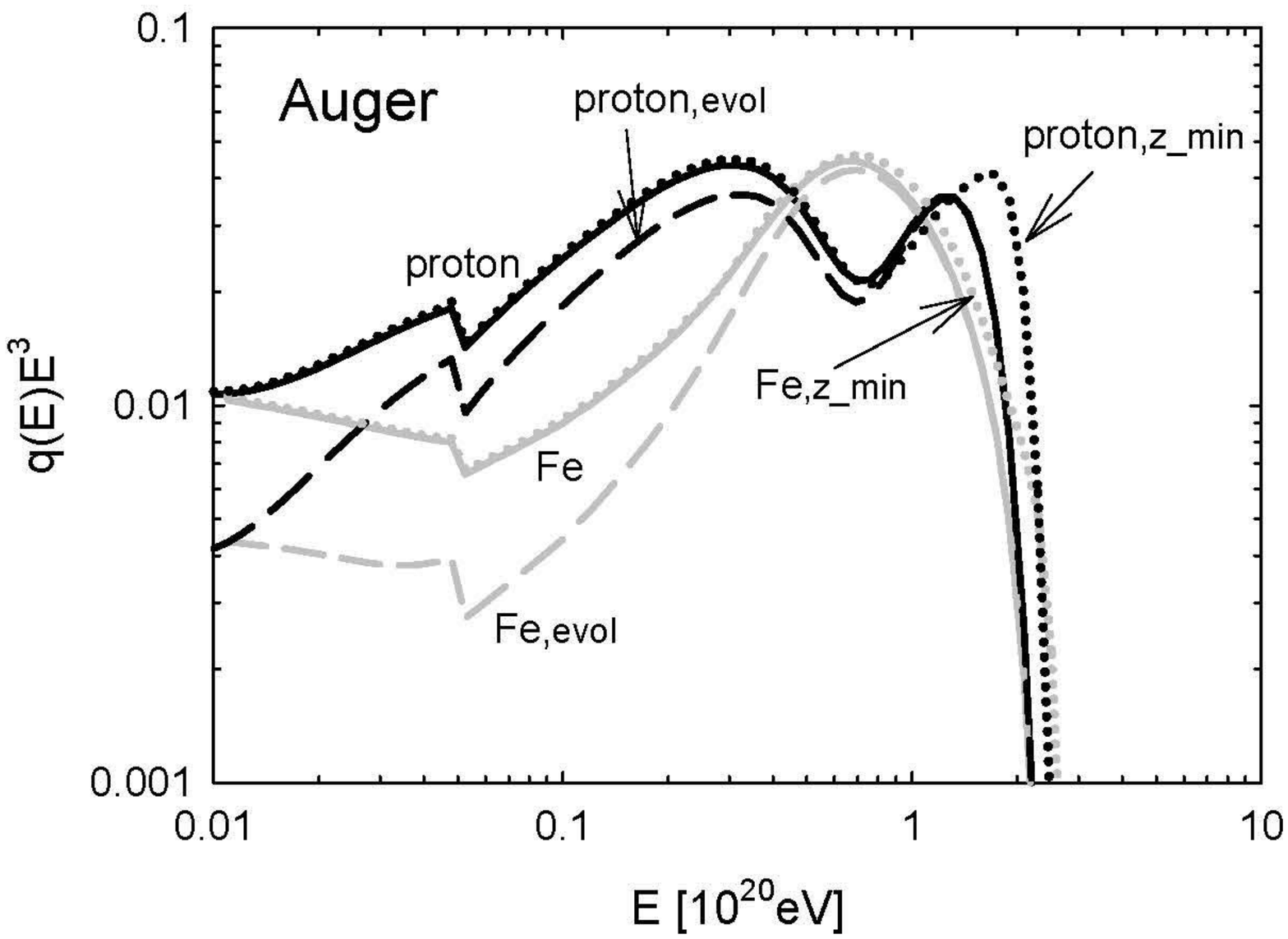}
\end{center}
\caption{Calculated source spectra in arbitrary units based on
the approximated analytically Auger data. Black lines for proton source;
gray lines for Iron source. Solid lines correspond to the homogeneous source
distribution without evolution, $m=0$. Dotted lines correspond to the spatial
source distribution with a finite distance to the nearest source located at
the redshift $z_{min} = 0.0024$ at $m=0$. Dash lines are the source spectra
for homogeneous source distribution with evolution described in the text.}
\label{sourceAUGER}
\end{figure}

\begin{figure}[tbp]
\begin{center}
\includegraphics[width=10.0cm]{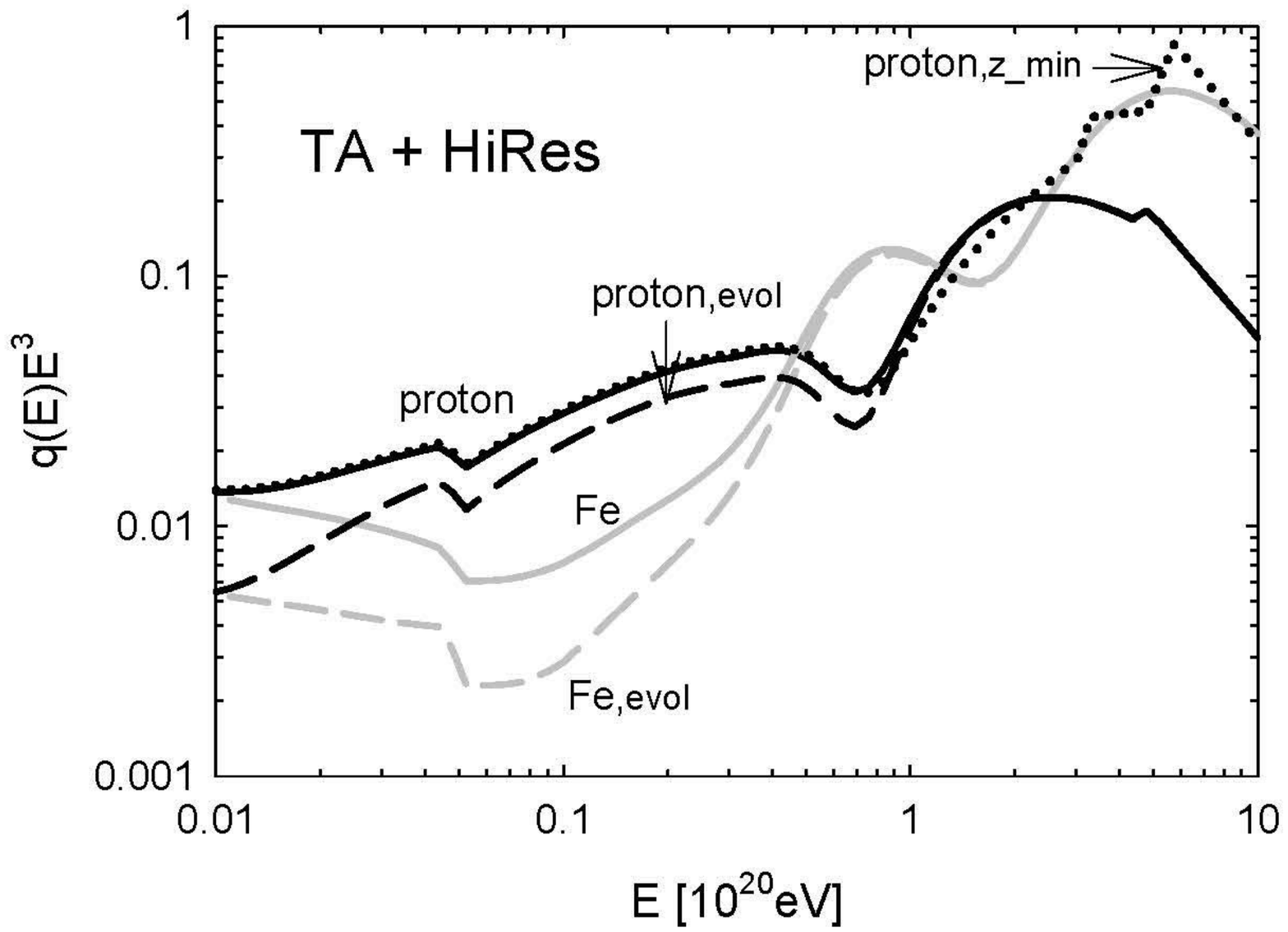}
\end{center}
\caption{The same as figure \ref{sourceAUGER} but for the approximated
analytically TA+HiRes data.}
\label{sourceTAHiRes}
\end{figure}

We first make calculations of the source spectra in two
simple cases of pure proton and pure Iron source composition. The results
are shown in figures \ref{sourceAUGER} for the Auger data and \ref{sourceTAHiRes}
for the TA+HiRes data. The dark lines refer to a pure proton source and the gray
lines refer to a pure Iron source. The solid lines illustrate the case without
source evolution $m=0$ (as it can be if the sources are the BL Lacs type
galaxies, see \cite{Berez06}); the dash lines describe the case of AGN with
a strong evolution where $m=3.2$ at $z<1.2$ and the evolution is saturated
at larger $z$ \cite{Berger05}. It is clear from
figures \ref{sourceAUGER} and \ref{sourceTAHiRes} that the strong cosmological
evolution leads to the decrease of required source power at low cosmic
ray energies. The difference with the source spectrum without evolution
reaches the factor of about $4$ at $10^{18}$ eV. The dotted lines show
the results of calculations with a non-zero distance to the nearest
source $z_{min}\neq 0$ at the source number density
$n_{s} = 10^{-4}$ $\textrm{Mpc}^{-3}$. It is clear that the finite
distance to the nearest source requires the increase of source power
at the highest energies of accelerated particles. The power law asymptotic
behavior of the TA+HiRes cosmic ray spectrum (\ref{eq:analytTA}) at the highest
energies can not be reproduced for the pure Iron source at
$n_{s} = 10^{-4}$ $\textrm{Mpc}^{-3}$.

The kinks in all source spectra at about $5\times 10^{18}$ eV are due to
the corresponding discontinuities of the first derivatives of the expressions
(\ref{eq:analytTA}) and (\ref{eq:analytAuger}) and in this sense they are artificial.
The power law tail at the highest energies in eq. (\ref{eq:analytTA}) describing
the observed spectrum results in the extension of the calculated TA+HiRes source
spectra to higher energies compared to the calculated Auger source spectra.
The difference between these two experiments is much smaller at energies below
$(3...5)\times 10^{19}$ eV where the data are more accurate.
The difference between proton and Iron sources at these relatively low energies
is evident: the proton source spectrum can be better approximated by a power law
compared to the concave Iron spectrum. It is explained by the influence of energy
loss on the $e^{-}e^{+}$ production in the case of ultra high energy protons
moving through the background radiation \cite{Berez06}. After propagation in
the intergalactic space, this process produces the characteristic dip in the
initial power law proton spectrum at around $5\times 10^{18}$ eV while the Iron
nuclei preserve the shape of their source spectrum.

\begin{figure}[tbp]
\begin{center}
\includegraphics[width=10.0cm]{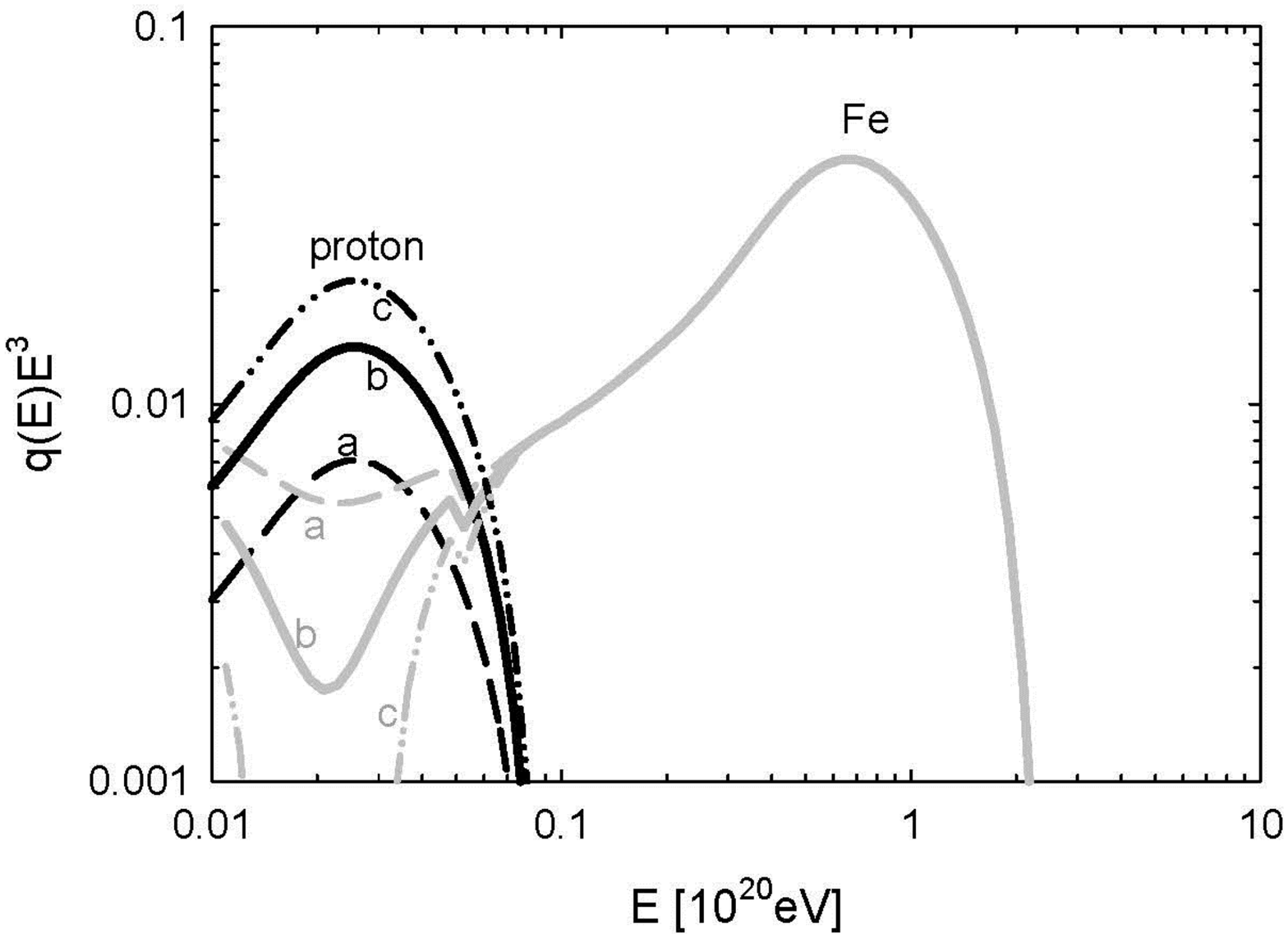}
\end{center}
\caption{Calculated source spectra based on
the Auger data for different Iron-to-proton source ratios indicated by
(a) for $S_{Fe}/S_{p}=2\times 10^{-2}$,
(b) for $S_{Fe}/S_{p}=10^{-2}$, and (c) for $S_{Fe}/S_{p}=6.7\times 10^{-3}$.
Black lines for proton
source; gray lines for Iron source. Proton and Iron source spectra have the same
dependence on magnetic rigidity. Homogeneous source distribution without evolution is assumed.}
\label{source2}
\end{figure}

The results of calculations when both proton and Iron are present in the source and
their spectra are similar functions of magnetic rigidity are shown in figure \ref{source2}.
One can see how the results of calculations depend on the assumed Iron-to-proton
source ratio $S_{Fe}/S_{p}$.

\begin{figure}[tbp]
\begin{center}
\includegraphics[width=10.0cm]{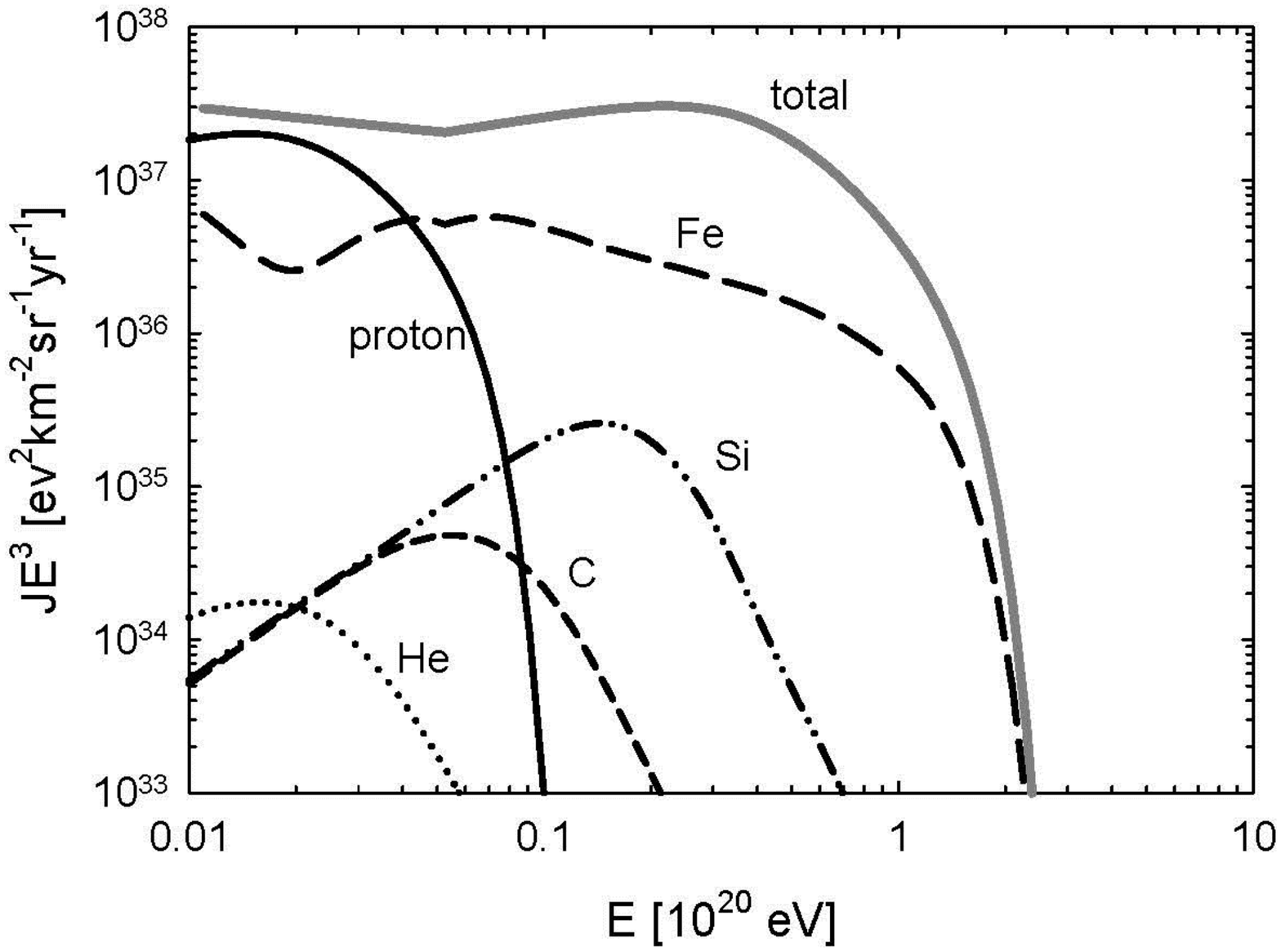}
\end{center}
\caption{Calculated spectra of different types of nuclei for Iron-to-proton
source ratio $S_{Fe}/S_{p}=10^{-2}$ and the total Auger cosmic ray spectrum at the Earth.}
\label{calspectra}
\end{figure}

\begin{figure}[tbp]
\begin{center}
\includegraphics[width=10.0cm]{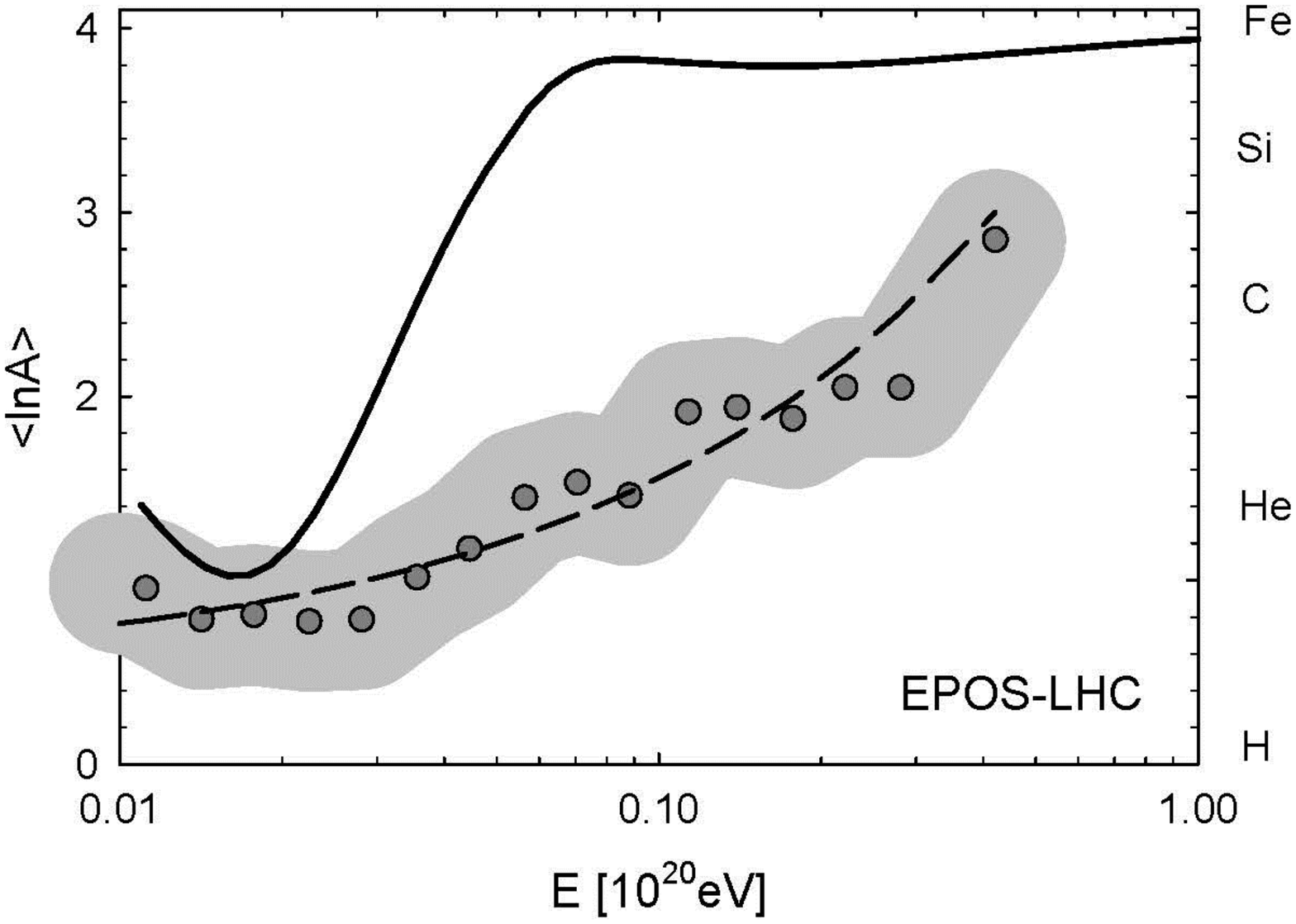}
\end{center}
\caption{Calculated value of $\langle ln(A) \rangle$ (solid line) together with
corresponding Auger data (dots and gray regions which characterizes errors in
determination of $\langle ln(A) \rangle$ in the EPOS LHC interaction model).
Dash line shows our approximation (\ref{lnA}).}
\label{composition}
\end{figure}

The calculated elemental composition of cosmic rays at the Earth for the
case $S_{Fe}/S_{p} = 10^{-2}$ is
presented in figure \ref{calspectra}. The corresponding value of $\langle ln(A) \rangle$
is shown in figure \ref{composition}. It is evident that our very simple model with
only two primary species at the source (protons and Iron nuclei), does not reproduce
the observed $\langle lnA \rangle$ except the energies $\sim 10^{18}$ eV where the
protons and light nuclei dominate and the highest energies where the Iron group nuclei dominate.
The intermediate nuclei are certainly needed at the source to reproduce
observations at all energies.

\section{The use of the measured mean logarithm of $A$}

The source spectra of protons and Iron can be found if the measurements of the mean
logarithm $\left< \ln A\right> $  are available in addition to the all-particle spectrum.
Then using Eq. (\ref{eq:greennum}) one has the following two systems of equations
for $q_j(1)$ and $q_j(56)$:

\begin{equation}
\sum_AN_{i}(A,z=0)=\sum_{A,j}(\triangle
\varepsilon)_{j}(G_{ij}(A;1)q_{j}(1)+56^{-1}G_{ij}(A;56)q_{j}(56)),
\label{eq:lnA1}
\end{equation}

\begin{equation}
\left< \ln A\right> _i\sum_AN_{i}(A,z=0)=\sum_{A,j}(\triangle
\varepsilon)_{j}G_{ij}(A;56)\frac {\ln (A)}{A}        q_{j}(56),
\label{eq:lnA2}
\end{equation}

The source spectrum of Iron $q_j(56)$ is found from the last system of equations. After that
 the source spectrum of protons $q_j(1)$ is obtained solving the system of equations (\ref{eq:lnA1}).

\begin{figure}[tbp]
\begin{center}
\includegraphics[width=10.0cm]{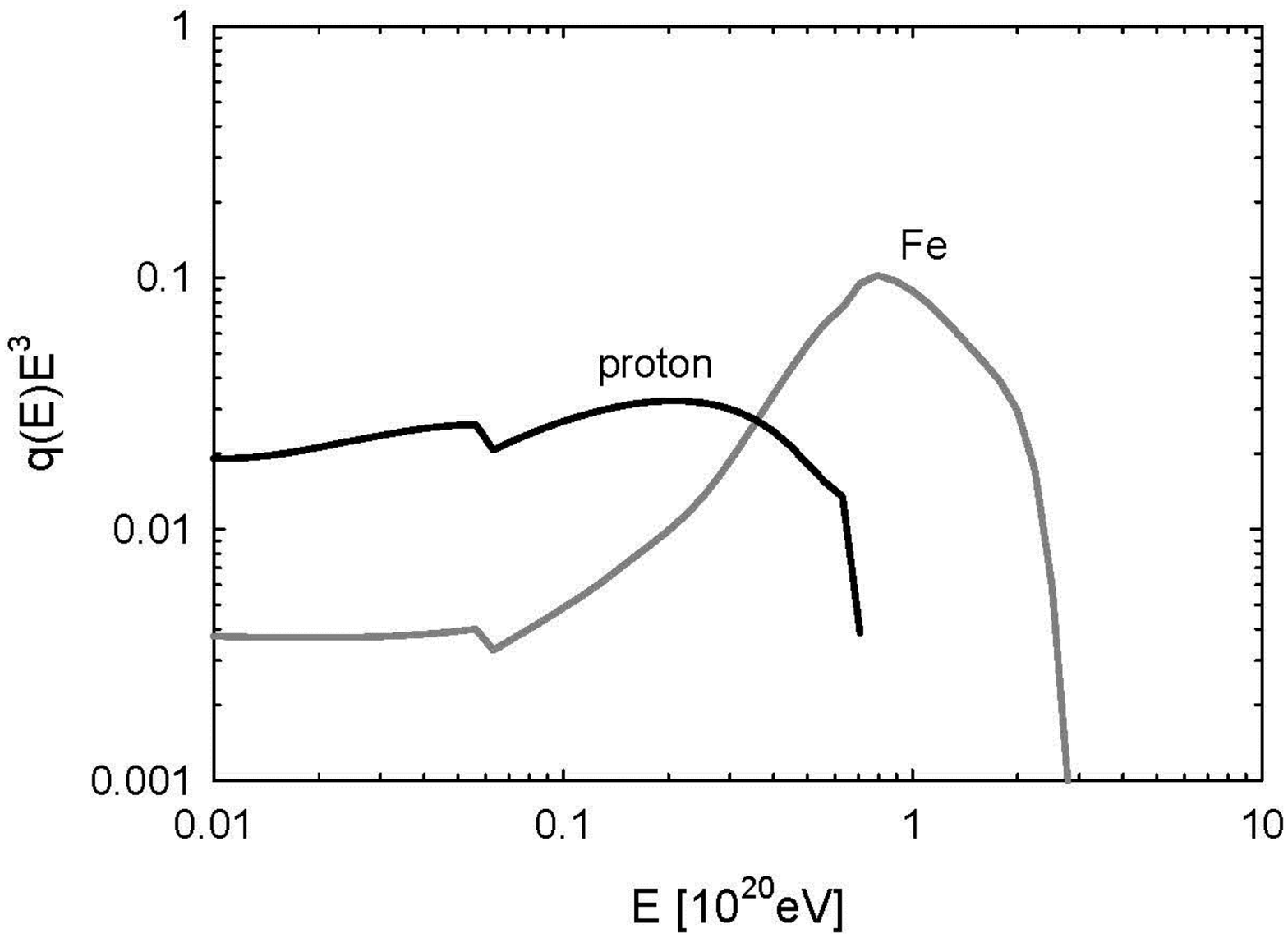}
\end{center}
\caption{Calculated source spectra of protons and Iron based on
Auger data on cosmic ray spectrum and $\langle lnA \rangle$.
Homogeneous source distribution without evolution is assumed.}
\label{onLnA}
\end{figure}

As a result both the observed all particle cosmic-ray Auger spectrum at the Earth approximated by
eq. (\ref{eq:analytAuger})
and the $\langle lnA(E) \rangle$ given by eq. (\ref{lnA}) can be exactly reproduced assuming that
only protons and Iron nuclei are present in the source. The corresponding calculated
source spectra are shown in figure \ref{onLnA}. They have different dependence on magnetic rigidity.

\section{Discussion and Conclusion}
\label{sec:conclusion}

We showed how one can find average spectrum of extragalactic sources
from the cosmic ray spectrum observed at the Earth. This task was formulated as an inverse
problem for the system of transport eqs. (\ref{eq:transport})
that describe the propagation of ultra-high energy cosmic rays in the
expanding Universe filled with the background electromagnetic
radiation.

The purpose of the present paper was the demonstration
of general approach to the problem. The simple settings were considered.
The two variants of observed cosmic ray spectrum were used in the calculations. One was taken
from the Auger data \cite{Auger13} and approximated by
formula (\ref{eq:analytAuger}). The other presented the combined TA+HiRes
data \cite{TAHiRes13} approximated by eq. (\ref{eq:analytTA}).
It was assumed that only two
kinds of nuclei (proton and Iron) were present in the sources and the cases of a
pure proton, pure Iron, and the mixed source composition were considered. In the last
case the calculations were made for the following two scenarios: 1) the proton and Iron source
spectra have the same shape on rigidity; 2) the source spectra and the Iron-to-proton
ratio are forced to reproduce the observed at the Earth value of $\langle lnA \rangle$.
Mathematically, the inverse problems for transport equations (\ref{eq:transport}) are
ill-posed in the general case that manifests itself in the instability of derived
solutions. It explains to a large extent simple assumptions used in the present paper.
We plan specifically study this problem in a future publication.

Two additional factors that may impact on the interpretation of the derived source
spectra at energies close to $10^{18}$ eV is worth to mention. The first is the possible
contribution of the Galactic sources that may dominate in the observed cosmic ray
spectrum at $< 3\times 10^{18}$ eV, see e.g. \cite{Berez06} for discussion. The second is
the strong deflection of cosmic ray trajectories in magnetic field that may produce the so called magnetic
 horizon effect in the expanding Universe. It is essential for the wide range of magnetic fields
 $0.1-10$ nG and distances between sources $d\geq 50$ Mpc at particle energies $E< 10^{18}Z$ eV
  \cite{AloisioBerez05}. The particle transmission factor at various
  distances to the source as a function of particle energy was calculated in \cite{KoteraLemoine08}.
 This factor characterises the suppression of cosmic ray intensity due to the magnetic
 horizon effect. Using results of these calculations and assuming that the value of the
 intergalactic magnetic field is $1$ nG and its correlation length is $1$ Mpc, one can
 find that the magnetic horizon effect is not significant for Iron nuclei with energies
 above $10^{18}$ eV if cosmic ray source density is not smaller than $\sim 10^{-4} \textrm{Mpc}^{-3}$.
 It is in the limits of the low bound on cosmic ray sources density found at the GZK energies
 in the Auger experiment \cite{AUGER2013}. It should be pointed out that the source
 density may increase with the decreasing of energy of accelerated particles. For example,
 such a behavior was found in the model of cosmic ray acceleration by the AGN jets with
 the observed distribution on kinetic energy where more numerous weak jets contribute most
 to small cosmic ray energies \cite{ASR}. The experimental
 indication of this effect was found in \cite{TakamiSato09}.
 The calculations of the source spectra in the present paper ignored the presence of the
 intergalactic magnetic field. It means that the found source spectra may contain
 depending on magnetic rigidity transmission factor that should be calculated
 independently assuming some magnetic field properties and the cosmic ray source distribution.

It is difficult to make firm astrophysical statements about cosmic ray source spectra and
composition from our simple modelling. However, some conclusions can be made.
Recall that the kinks in the calculated source spectra at about $5\times 10^{18}$ eV
reflect the corresponding discontinuities in the derivatives of the approximation
eqs. (\ref{eq:analytTA}, \ref{eq:analytAuger}) and are artificial in this sense. To
demonstrate the specific character of the inverse problem solutions, we did
not correct the unphysical approximations of the observed spectra.

Accepting that the TA+HiRes data favour the proton source, one can see
from figure \ref{sourceTAHiRes} that the source spectrum at $m=0$ is close to the
power law $\propto E^{-2.6}$ below $\sim 4\times 10^{19}$ eV with some deviations above
this energy and bending down at $\gtrsim 5\times10^{20}$ eV. Notice that statistics above
$3\times 10^{19}$ eV is poor and the power law approximation of data (\ref{eq:analytTA})
is not very reliable at the highest energies. The proton source spectrum $E^{-2.55}$
to $E^{-2.75}$ was derived in the solution of direct transport problem under the
assumption of a power-law source spectrum without cosmological
source evolution \cite{Berez06}.

The Auger data favor the transition from a proton source composition to the Iron one
as the energy is rising. With our simple two-species composition, this case is most
closely reproduced by the calculations illustrated in figure \ref{source2}. The
obtained source spectra resemble the results \cite{Allard08,Aloisio09}
based on the analysis of direct transport problems with a power law source spectrum.
The maximum magnetic rigidity of accelerated particles $(3...5)\times 10^{18}$ eV is
relatively low in this case that alleviates the problem of cosmic ray acceleration to the extremely high energies.
The calculated composition of cosmic rays at the Earth shown in figure \ref{composition}
considerably deviates from the Auger measurements and certainly requires incorporation
of the intermediate nuclei between protons and Iron in the source composition.

The study of inverse transport problem is a useful tool for the investigation of ultra
high energy cosmic rays allowing the abandonment of the standard assumption of power law
source spectrum with an abrupt cutoff at some maximum magnetic rigidity as it is usually
assumed when the direct problem is considered. The present work is a simple
illustration of this new approach.


\acknowledgments

We thank Anatoly Lagutin for useful remarks. The work was
supported by the Russian Foundation for Basic Research grant
13-02-00056 and by the Russian Federation Ministry of Science and
Education contract 14.518.11.7046. The work was partly fulfilled
during VSP visit to the University of Maryland, USA, where it was
supported by the NASA grant NNX13AC46G.


\end{document}